\documentclass[preprint]{aastex62}

\usepackage{amsmath}
\usepackage{graphicx}% Include figure files
\newcommand{\ps}{\ {\rm s}^{-1}}

\newcommand{\cm}{\ {\rm cm}}
\newcommand{\km}{\ {\rm km}}

\newcommand{\TeV}{\ {\rm TeV}}
\newcommand{\uG}{\ \mu{\rm G}}

\newcommand{\gray}{$\gamma$-ray}
\newcommand{\grays}{$\gamma$-rays}

\newcommand{\Emaxc}{E_{\rm max}^{\rm cool}}
\newcommand{\Emaxa}{E_{\rm max}^{\rm age}}
\newcommand{\Ebre}{E_{\rm bre}}
\newcommand{\Ecut}{E_{\rm cut}}
\newcommand{\Emax}{E_{\rm max}}
\newcommand{\tsed}{t_{\rm sed}}

\newcommand{\rosat}{{\sl ROSAT}}
\newcommand{\hess}{{\rm H.E.S.S.}}
\newcommand{\Fermi}{{\sl Fermi}}

\newcommand{\suzaku}{{\sl Suzaku}}
\received{xxx}
\revised{xx.xx}
\accepted{xx.xx}

\submitjournal{xxx}

\begin{document}

%% Title
\title{Electron Acceleration in Middle-age Shell-type $\gamma$-Ray Supernova Remnants}

%% Authors
\email{liusm@pmo.ac.cn; xiaozhang@nju.edu.cn}

\author[0000-0002-9392-547X]{Xiao Zhang}
\affil{School of Astronomy \& Space Science, Nanjing University, 163 Xianlin Avenue, Nanjing~210023, China}
\affil{Key Laboratory of Modern Astronomy and Astrophysics, Nanjing University, Ministry of Education, Nanjing~210023, China}

\author[0000-0003-1039-9521]{Siming Liu}
\affil{Key Laboratory of Dark Matter and Space Astronomy, Purple Mountain Observatory, Chinese Academy of Sciences, 8 Yuanhua Road, Nanjing 210034, China}

\begin{abstract}

Over the past decade, \gray\ observations of supernova remnants (SNRs) and accurate cosmic-ray (CR) spectral measurements have significantly advanced our understanding of particle acceleration in SNRs.
In combination with multiwavelength observations of a large sample of SNRs, it has been proposed that the highest energy particles are mostly accelerated in young remnants, and the maximum energy that middle-age and old SNRs can accelerate particles to decreases rapidly with the decrease in shock speed. 
If SNRs dominate the CR flux observed at Earth, a large number of particles need to be accelerated in old SNRs for the soft CR spectrum even though they cannot produce very high-energy CRs. 
With radio, X-ray, and \gray\ observations of seven middle-age shell-type SNRs, we derive the distribution of high-energy electrons trapped in these remnants via a simple one-zone leptonic emission model and find that their spectral evolution is consistent with such a scenario. 
In particular, we find that particle acceleration by shocks in middle-age SNRs with an age of $t$ can be described by a unified model with the maximum energy decreasing as $t^{-3.1}$ and the number of GeV electrons increasing as $t^{2.5}$ in the absence of escape from SNRs.
\end{abstract}

%\titlerunning{}
%\maketitle

%%%%%%%%%%%%%%%%%%%%%%%%%%%%%%%%%%%%%%%%%%%%%%%%%%%%%%%%%%%%%%%%%%%%%%%%%%%%%%%%
%%%%%%%%%%%%%%%%%%%%%%%%%%%%%%%%%%%%%%%%%%%%%%%%%%%%%%%%%%%%%%%%%%%%%%%%%%%%%%%%
%%%%%%%%%%%%%%%%%%%%%%%%%%%%%%%%%%%%%%%%%%%%%%%%%%%%%%%%%%%%%%%%%%%%%%%%%%%%%%%%
\section{Introduction}
Since it was first proposed that supernovae may be responsible for the production of cosmic rays (CRs) in the 1930s \citep{Baade1934b}, it has been widely accepted that CRs up to the spectral knee energy of $\sim3\times10^{15}$ eV mainly originate in the Milky Way, presumably being accelerated by shocks of supernova remnants (SNRs).
Now we know that energetic electrons and protons can indeed be accelerated in the shock waves of SNRs. 
Evidence for acceleration of electrons first comes from the SNRs' radio continuum spectra, which are produced by relativistic electrons through the synchrotron process.
Moreover, the detection of nonthermal X-ray emission in the SNR SN~1006 shows that the SNR shocks can boost electrons up to TeV energies \citep{Koyama1995}.
With regard to protons, evidence of their acceleration in SNRs was confirmed recently by the discovery of the pion-decay \gray\ signal in SNR IC~443 \citep{Ackermann2013}, W44 \citep{W44.AGILE.2011,Ackermann2013,W44.AGILE.2014} and W51C \citep{W51C.Fermi.2016}.

The presence of both energetic electrons and protons in SNRs makes the origin of their observed \gray\ emission ambiguous, because both leptonic and hadronic processes can generate \gray\ emission efficiently.
In the leptonic scenario, the \grays\ are produced via inverse Compton (IC) scattering on the background photon field by relativistic electrons and/or via the bremsstrahlung process.
In the hadronic models, the \grays\ come from the decay of $\pi^{0}$ produced mainly through inelastic collisions between relativistic protons and ambient nuclei.
Based on these two scenarios, various models have been proposed to explain the observed \gray\ emissions from individual sources \citep[e.g.,][]{Li2010.W28,Uchiyama2010,Ellison2011,Ohira2011,Tang2011.model,Atoyan2012,Morlino2012.tycho,Zhang2013.tycho,Berezhko2013.tycho,Gabici2014,Tang2014.C,Zhang2016.J1713,Ohira2017}.

With the development of the ground-based \gray\ telescopes like \hess, VERTIAS, MAGIC etc, about two dozen TeV-bright SNRs\footnote{http://tevcat.uchicago.edu/} have been detected.
With the accumulation of data from the \Fermi\ satellite, the first \Fermi-LAT SNR catalog with identified SNRs and candidates has been built.
These observations reveal diversified \gray\ spectra, which \citet{Yuan2012.LB} attributed to variations in the density of the emission region.
The increasing number of \gray-bright SNRs also makes it possible to perform a population study \citep{Mandelartz2015,Zeng2018.snrs}.

Traditionally, the exploration of the SNR origin of galactic CRs has been focused on the maximum energy to which shocks of SNRs can accelerate particles \citep{Lagage1983}. 
Although it has been well recognized that strong shocks of young SNRs likely dominate the acceleration of the highest-energy CRs \citep{Bell2004}, the nondetection of SNRs above 100 TeV challenges SNRs as PeVatrons. 
Moreover, recent deep TeV observations of the brightest TeV SNR RX J1713.7$-$3946 reveal a convex \gray\ spectrum, which can be attributed to broken power law distribution of energetic particles not anticipated by most particle acceleration models. 
\citet{Ohira2017} first proposed that such broken power law particle distribution may be attributed to the time integration of accelerated particles by a gradually weakening shock that accelerates more less-energetic particles as the shock slows down.
Such a scenario is in line with a time-dependent particle acceleration model proposed by \citet{Zhang2017.LY} for the anomalous energy distribution of CR ions.

Using a simple one-zone emission model, \citet{Zeng2018.snrs} fit the spectral energy distribution of nonthermal emission from a sample of 35 middle-age or old SNRs. 
They found that in general, the particle distribution in these SNRs can be described by a broken power-law function with a high-energy cutoff and as the SNR ages, the break energy decreases and the low-energy spectrum becomes harder, in agreement with the model proposed by \citet{Ohira2017} and implying very efficient acceleration of low-energy CRs in old SNRs \citep{Zhang2018.Acc}.  
In this paper, we focus on the spectral modeling of seven shell-type bright \gray\ SNRs, namely SN~1006, RX~J1713.7$-$3946, RCW~86, RX J0852.0$-$4622, HESS~J1731$-$347, G150.3$+$4.5 and HESS~J1534$-$571. 
The basic properties of these SNRs are listed in Table~\ref{tab:samp}. 
All of them have relatively harder GeV spectra and have spatially well-correlated nonthermal emissions from radio to $\gamma$-rays so that a simple one-zone leptonic emission model is applicable, and in general the a high-energy electron distribution is assumed to follow a broken power law with a high energy cutoff.
The model is described in Section 2. In Section 3, we discuss spectral fits to each SNR. A unified model is proposed in Section 4 for the evolution of high-energy electron distribution in these SNRs. We draw conclusions in Section 5.

%%%%%%%%%%%%%%%%%%%%%%%%%%%%%%%%%%%%%%%%%%%%%%%%%%%%%%%%%%%%%%%%%%%%%%%%%%%%%%%%
%%%%%%%%%%%%%%%%%%%%%%%%%%%%%%%%%%%%%%%%%%%%%%%%%%%%%%%%%%%%%%%%%%%%%%%%%%%%%%%%
%%%%%%%%%%%%%%%%%%%%%%%%%%%%%%%%%%%%%%%%%%%%%%%%%%%%%%%%%%%%%%%%%%%%%%%%%%%%%%%%
\section{Modeling}

To study the overall spectral characteristics, we adopt a one-zone emission model.
In this model, electrons accelerated at the shock front obey a single power-law energy distribution with a maximum energy $\Emax$, and are injected into the main emission region.
The maximum energy is determined by the balance of acceleration and energy loss in early stages of the SNR shock evolution.
The acceleration time scale is given by $\tau_{acc}=\eta_{acc} D(E,t)/u_s^2$, where $\eta_{acc}\approx10$ is a numerical factor, $u_s$ is the shock speed, and $D(E,t)$ is the diffusion coefficient \citep{Drury1983}.
Following \citet{Ohira2012}, we assume that the diffusion coefficient has a Bohm-like form $D(E,t)=\eta_g cE/3eB(t)$, where $c$, $e$, $E$, and $B(t)$ are the speed of light, the elementary charge, the particle energy, and the magnetic field, respectively.
The gyrofactor $\eta_g$ in the diffusion coefficient is still not well understood, and maybe changing with the SNR age \citep{Aharonian2017}.
In this paper, we assume it has a power-law behavior $\eta_g\propto t^{\lambda}$ \citep[e.g.,][and references therein]{Ohira2012}.
Due to the dominance of synchrotron loss, the cooling time is given by $\tau_{syn}=1.25\times10^4\, (E/1\ {\rm TeV})^{-1}(B_{\rm SNR}/1 \uG)^{-2}$ kyr.
By setting $\tau_{acc}=\tau_{syn}$, we can get the cooling-limited maximum energy $\Emaxc$.
In very early stage of SNRs, the energy loss time is longer than the SNR age $t$ and the maximum energy is constrained by $t$, which gives an age-limited maximum energy $\Emaxa$.
Therefore, the maximum energy up to which relativistic electrons can be accelerated at the shock front is given by $\Emax={\rm min}(\Emaxc,\ \Emaxa)$, where $\Emaxc$ and $\Emaxa$ are given, respectively, by
\begin{eqnarray}
\Emaxc\approx110\,\left( \frac{\eta_g}{1} \right)^{-1/2}
            \left( \frac{B_{\rm SNR}}{1\uG}   \right)^{-1/2}
            \left( \frac{u_s}{10^8\cm\ps} \right) \TeV   \label{eq:emax_cool} \\
\Emaxa\approx1\,  \left( \frac{\eta_g}{1} \right)^{-1}
            \left( \frac{B_{\rm SNR}}{1\uG}   \right)
            \left( \frac{u_s}{10^8\cm\ps} \right)^2 
            \left( \frac{t}{1\ {\rm kyr}} \right) \TeV   \label{eq:emax_age}
\end{eqnarray}

Electrons injected into the emission region are subject to radiative energy loss. The accumulated electron distribution may be described with a broken power-law function. 
In the Sedov phase of SNRs, shock speed decreases, and so does the $\Emaxa$. In combination with an efficient injection, the time-integrated electron distribution may also be described by a broken power-law function \citep{Ohira2017}.

Without loss of generality, we assume that the time-integrated electron spectrum in the emission zone has a broken power-law form with a high-energy cutoff:
\begin{equation}
\label{eq:dis}
\frac{dN(E)}{dE} = A \cdot E^{-\alpha}
              \left[ 1+\left( \frac{E}{\Ebre} \right)^{\sigma}  \right]^{-\Delta\alpha/\sigma} {\rm exp}
              \left[  -\left( \frac{E}{\Ecut} \right)^{\beta}  \right]
\end{equation}
where $\Ebre$ is the break energy, $\alpha$ is the index below the break, $\beta$ characterizes the cutoff shape, $\sigma$ describes the smoothness of the transition at the break energy with $\sigma=5$ adopted here, and $\Delta\alpha$ is the index change.
The radio and X-ray emissions are produced by energetic electrons via synchrotron process while the $\gamma$-rays are produced via IC.
We adopt the photon emissivity given in \citet{Blumenthal1970} for synchrotron and IC processes.
Both the hadronic and bremsstrahlung processes are ignored because the density of the emission region for these seven sources is very low.

%%%%%%%%%%%%%%%%%%%%%%%%%%%%%%%%%%%%%%%%%%%%%%%%%%%%%%%%%%%%%%%%%%%%%%%%%%%%%%%%
%%%%%%%%%%%%%%%%%%%%%%%%%%%%%%%%%%%%%%%%%%%%%%%%%%%%%%%%%%%%%%%%%%%%%%%%%%%%%%%%
%%%%%%%%%%%%%%%%%%%%%%%%%%%%%%%%%%%%%%%%%%%%%%%%%%%%%%%%%%%%%%%%%%%%%%%%%%%%%%%%
\section{Results}

To calculate the IC emission, besides the microwave background radiation, we include an IR photon background with a temperature of 35 K and an energy density of 0.7 ${\rm eV\ cm^{-3}}$ except specified otherwise below. In general, there are seven model parameters that need to be determined by the spectral fit: $\alpha$, $\Delta\alpha$, $\Ebre$, $\Ecut$, $\beta$, the total energy of electrons above 1 GeV $W_e$, which determines the normalization of the electron distribution, and $B_{\rm SNR}$.
To facilitate the comparison of these SNRs, we assume $\beta = 2$ and $\Delta\alpha = 0.8$ except for SN~1006, which is the youngest SNR in our sample.
Justification for these assumptions will be given in Section 4. There are therefore five parameters.

\subsection{SN~1006 (G327.6+14.6)}
\begin{figure}[htb]
\centering
\includegraphics[height=50mm,angle=0]{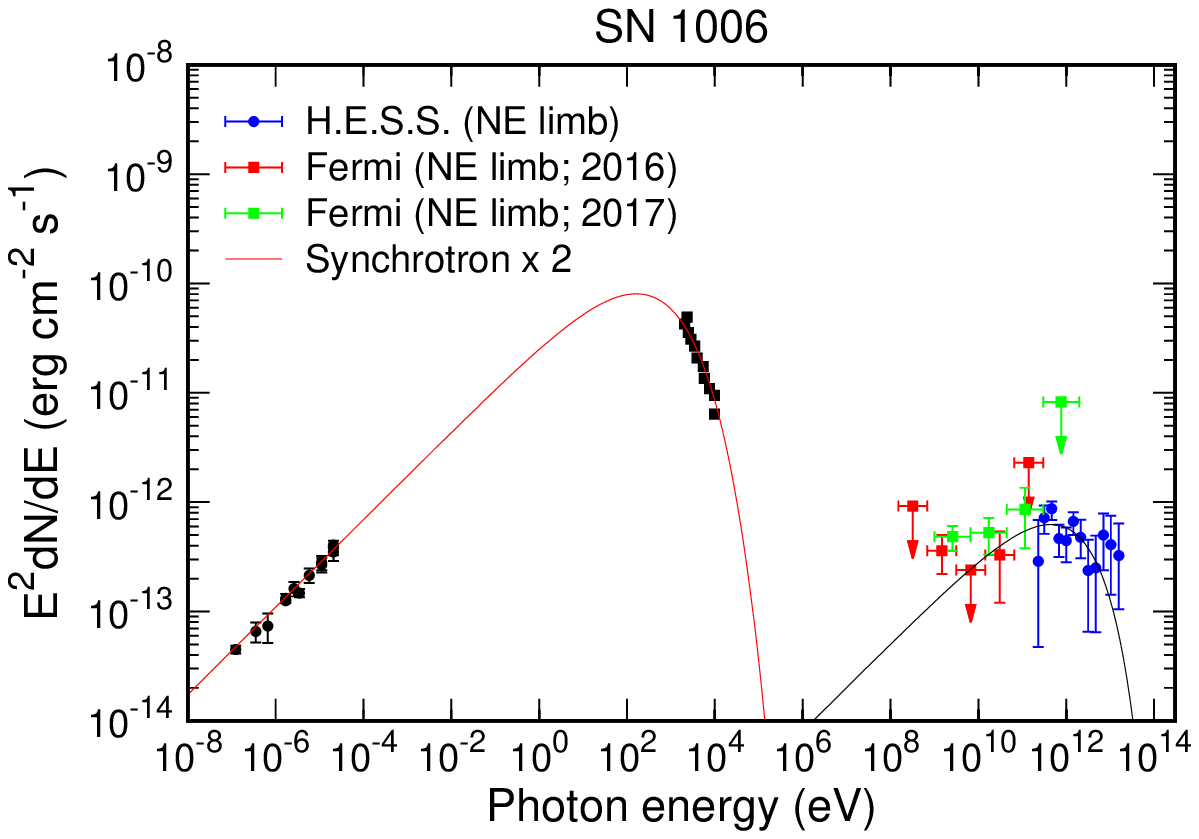}
\includegraphics[height=50mm,angle=0]{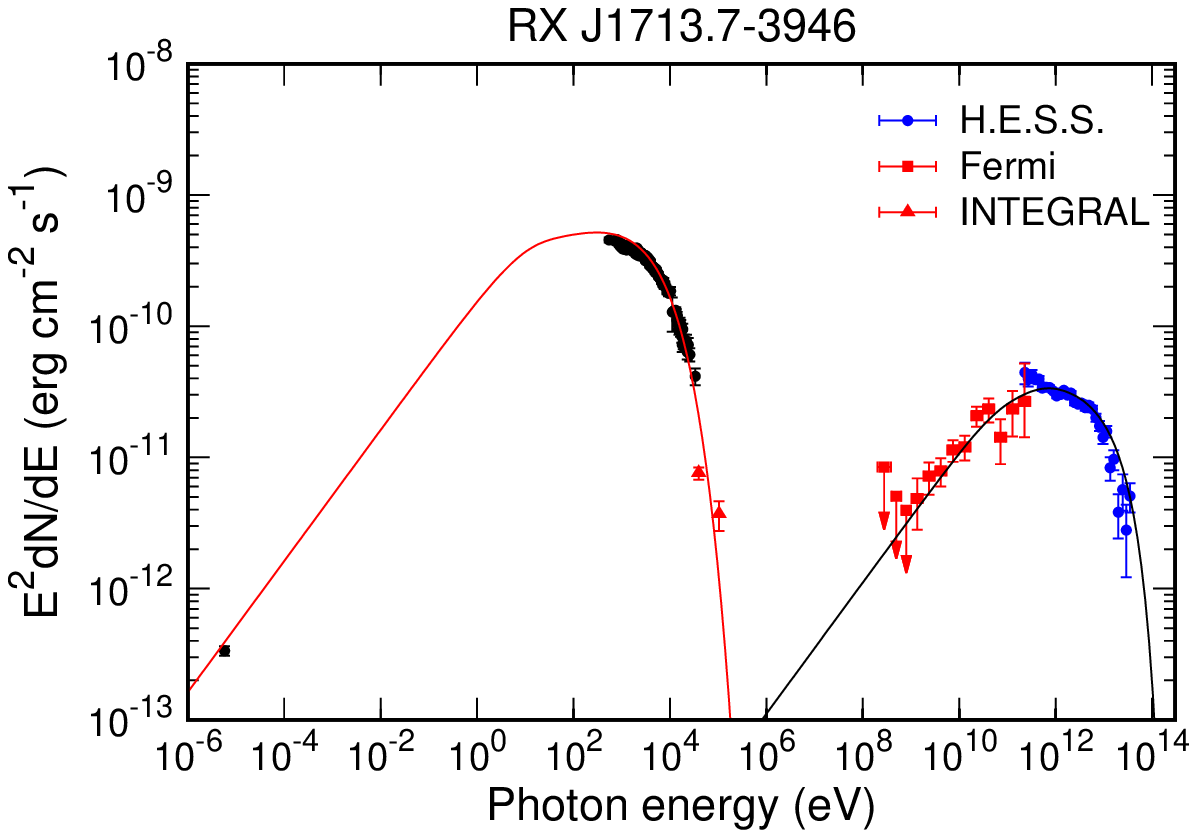}
\includegraphics[height=50mm,angle=0]{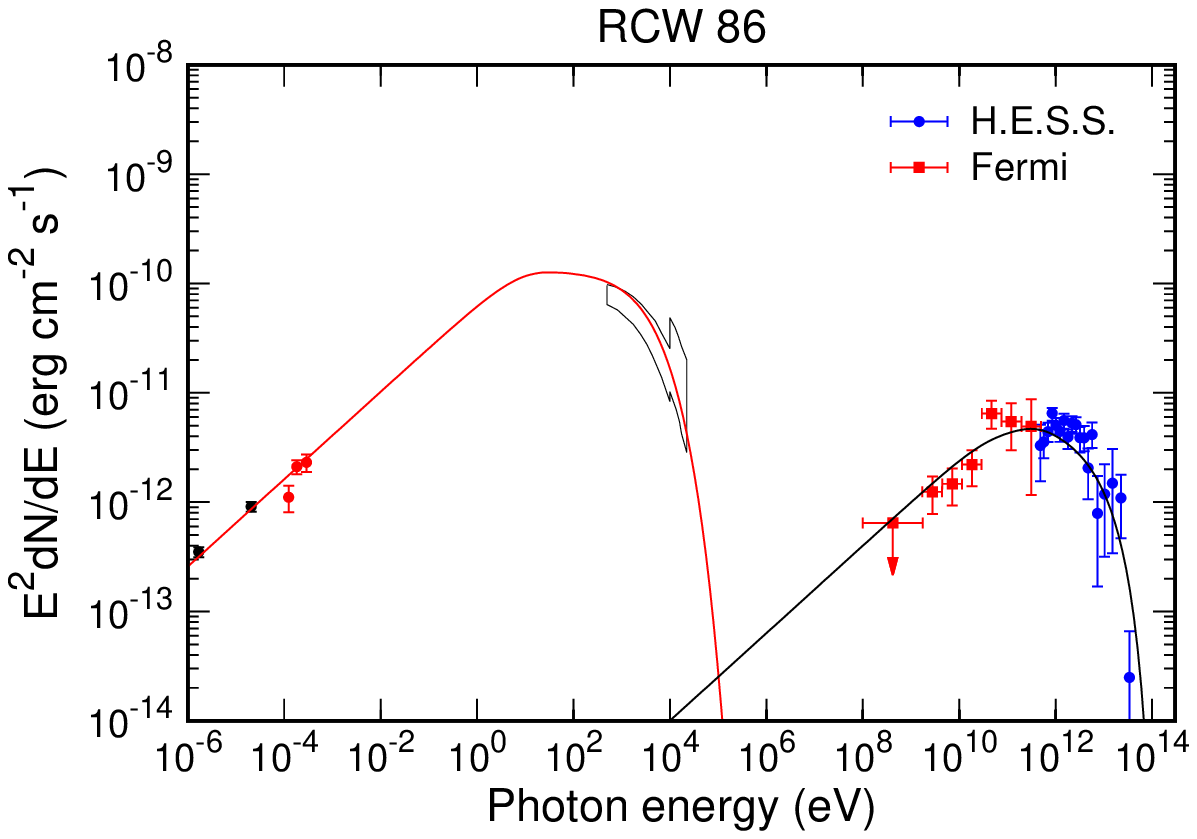}
\includegraphics[height=50mm,angle=0]{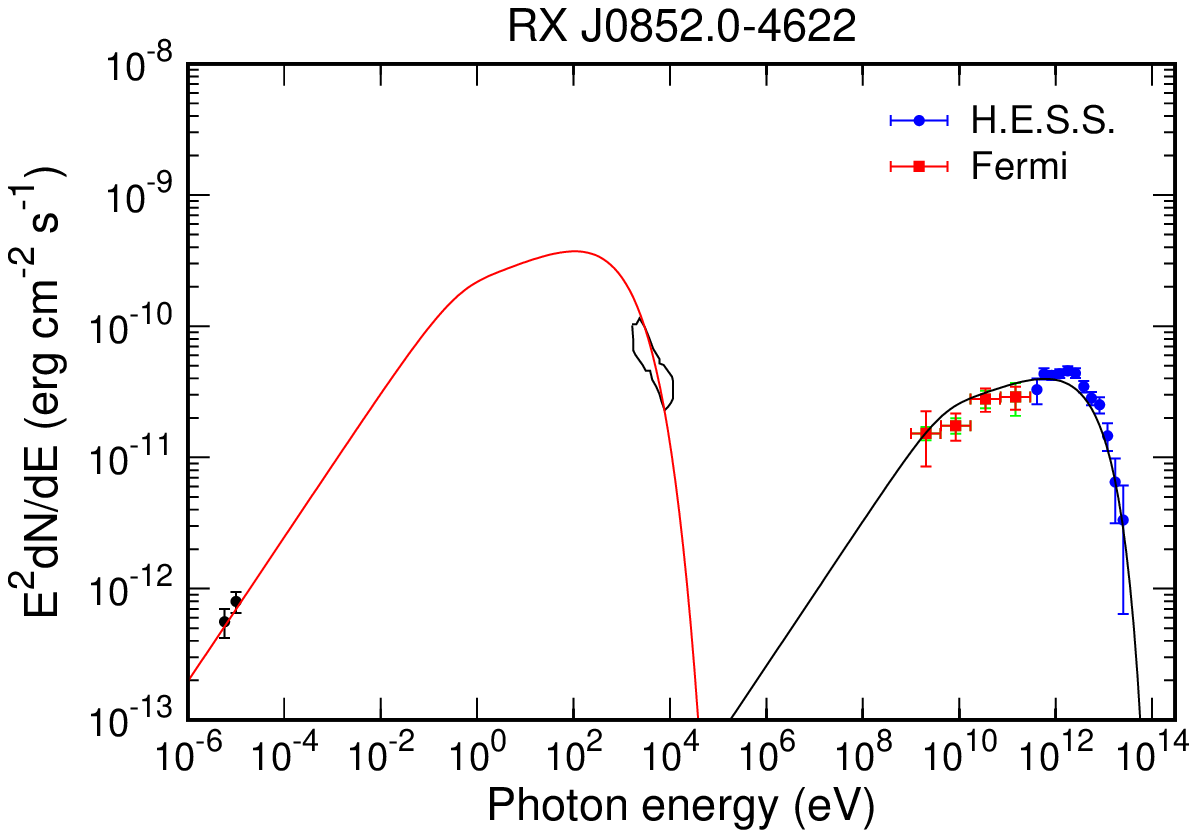}
\includegraphics[height=50mm,angle=0]{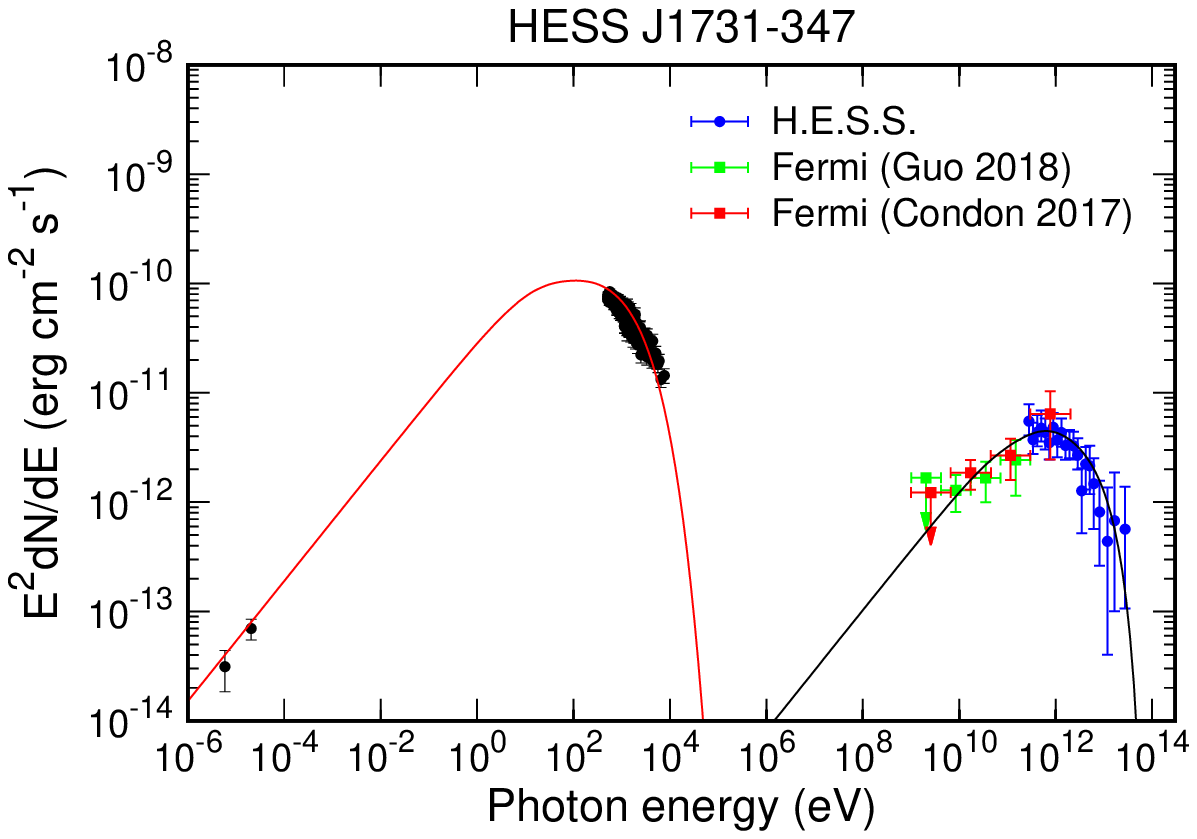}
\includegraphics[height=50mm,angle=0]{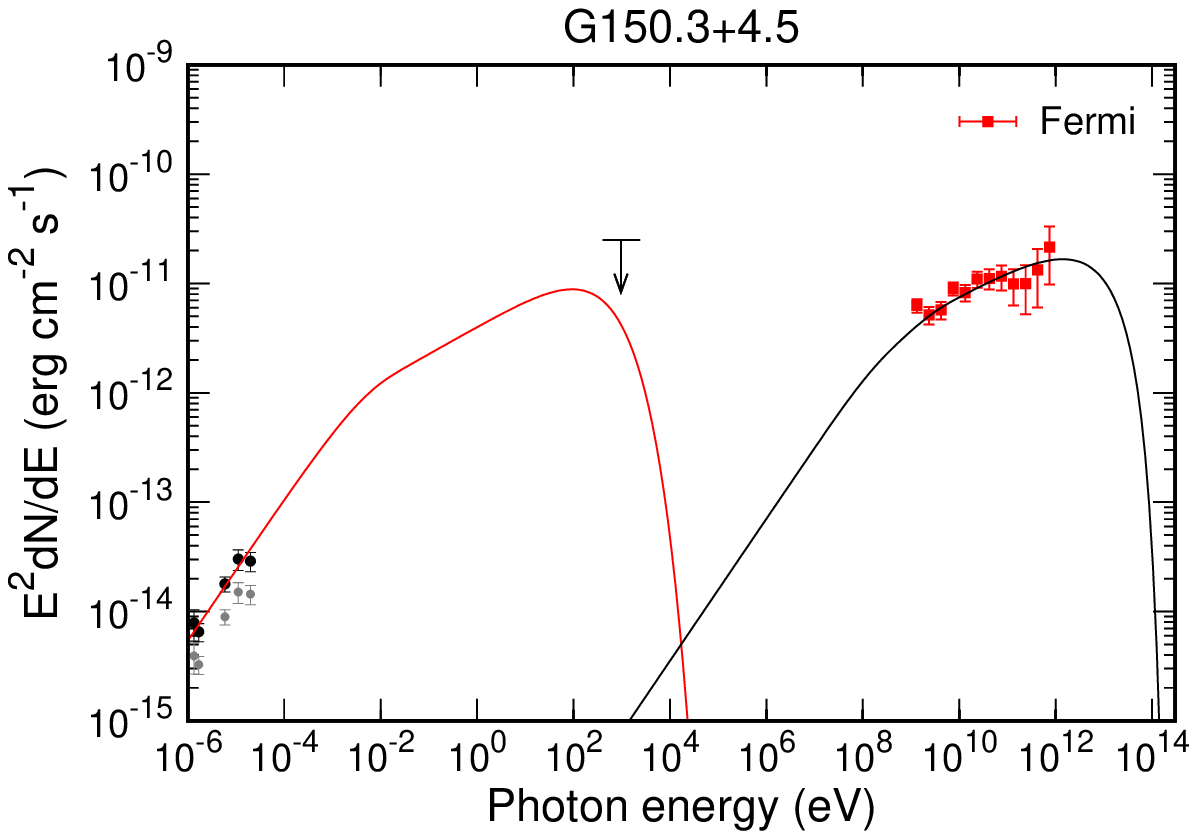}
\includegraphics[height=50mm,angle=0]{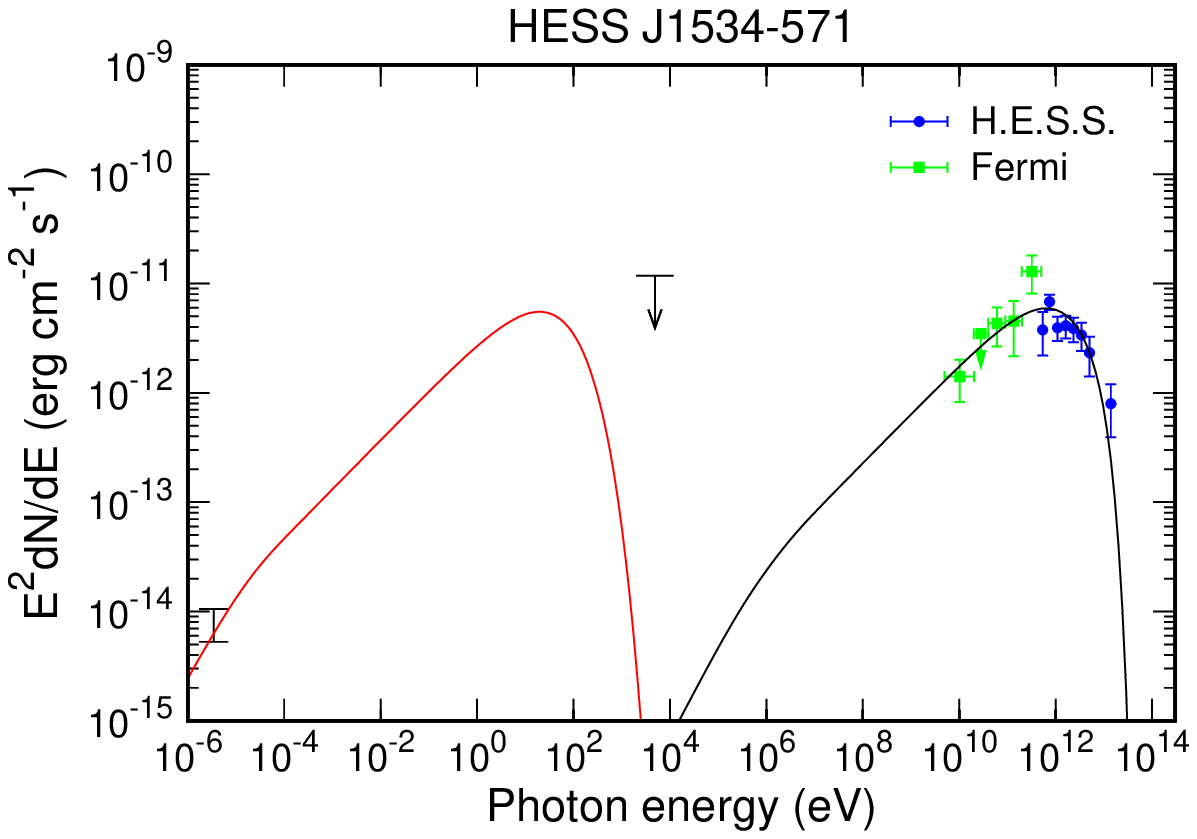}
\caption{Fit of the nonthermal spectrum of the seven SNRs.
References of the observational data ---
SN~1006: 
    radio    \citep[][and references therein]{Allen2001}, 
    \Fermi\  \citep{SN1006.Fermi.2016,SN1006.J1731.Fermi.2017}, 
    X-ray and \hess\  \citep{SN1006.HESS.2010};
RX~J1713.7$-$3946: 
    radio    \citep{Acero2009}, 
    X-ray    \citep{Tanaka2008},
    INTEGRAL \citep{4th.IBIS.2010},
    \Fermi\ and \hess\ \citep{J1713.HESS.2018};
RCW~86: 
    radio    \citep{Caswell1975}, 
    X-ray    \citep{RCW86.Fermi.2012},
    \Fermi\  \citep{RCW86.Fermi.2016}
    \hess\   \citep{RCW86.HESS.2018};
RX~J0852.0$-$4622: 
    radio    \citep{Duncan2000}, 
    \Fermi\  \citep{J0852.Fermi.2011}, 
    X-ray    \citep{J0852.HESS.2007},
    \hess\   \citep{J0852.HESS.2018};
HESS~J1731$-$347: 
    radio    \citep{Tian2008.J1731}, 
    \Fermi\  \citep{SN1006.J1731.Fermi.2017,J1731.Fermi.2018}, 
    X-ray    \citep{Doroshenko2017},
    \hess\   \citep{J1731.HESS.2011};
G150.3+4.5:
    radio    \citep{Gerbrandt2014},
    X-ray and \Fermi\ \citep{Cohen2016.PhDT};
HESS J1534$-$571:
    \Fermi\  \citep{J1534.Fermi.2017},
    radio, X-ray and \hess\ \citep{newSNR.HESS.2018}.
}
\label{fig:sed}
\end{figure}

% SN~1006 (G327.6+14.6) \quad
SN 1006 is the remnant of a historical supernova that exploded on AD 1006 May 1 \citep{Stephenson2002}.
It is a typical bilateral SNR with most nonthermal emissions originating from two large segments on opposite sides of the remnant.
In the GeV band, only the northeast (NE) shell was detected by \Fermi-LAT \citep{SN1006.Fermi.2016,SN1006.J1731.Fermi.2017}, showing a hard GeV spectrum.
Due to lack of GeV data for the southwest shell, we only fit \gray\ data of NE shell and the synchrotron spectrum is multiplied by a factor of 2 to match the radio and X-ray fluxes of the whole remnant. 
We find that a single power-law distribution with a high-energy cutoff can fit the nonthermal emission spectrum and for $\beta=1$, the energy loss time at the cutoff energy due to synchrotron radiation is comparable to the age of the remnant.
The corresponding model parameters are given in Table 2.
The spectral fit is shown in Figure~\ref{fig:sed}.

\subsection{RX~J1713.7$-$3946 (G347.3--0.5)}% \quad
% RX~J1713.7$-$3946 (G347.3--0.5) \quad
RX~J1713.7$-$3946 is a Galactic SNR discovered in X-rays by {\sl ROSAT} \citep{Pfeffermann1996} and has an accurate age measurement due to its association with the historical SN AD~393 \citep{Wang1997.QC}.
It has extremely faint radio emission and prominent nonthermal X-ray emission, both showing shell morphology with enhancement in the northwest portion.
Observations of the H.E.S.S. experiment first resolved TeV \gray{s} from the shell, and its ringlike shape closely matches the nonthermal X-ray emission \citep{J1713.HESS.2007}.
The GeV \gray\ spectrum has a photon index of 1.5 \citep{J1713.Fermi.2011} and can be smoothly connected with TeV data. 
The GeV-TeV combination analyses performed by the \cite{J1713.HESS.2018} with deeper observations show that there is a spectral break around 100 GeV. 
For $\Delta\alpha=0.8$, $\beta=2.0$, and infrared photons with a temperature of 26.5 K and an energy density of 0.42 eV~cm$^{-3}$ \citep{Finke2012}, we obtain the following best-fit model parameters: $\alpha=2.0$ $W_e=4.0\times10^{47}$ erg, $B_{\rm SNR}=16\ \uG$, $\Ebre=3.5$ TeV, and $\Ecut=80$ TeV.

\subsection{RCW~86 (G315.4$-$2.3)}
% RCW~86 (G315.4$-$2.3) \quad
RCW~86 is a radio shell-type Galactic SNR with a nearly circular shape and the diameter is about $42^{\prime}$ and the radio spectral index is about 0.6 \citep{Caswell1975,Kesteven1987}.
Although the inferred young age of RCW~86 seems to be inconsistent with the large physical size at a distance of $\sim2.5$ kpc \citep{Rosado1996, Sollerman2003,Helder2013}, it is likely the remnant of the first historical SN seen by the Chinese in AD 185 \citep[e.g.,][]{Vink2006}.
At the GeV band, only upper limits were derived toward this source by \citet{RCW86.Fermi.2012}, and \gray\ emission from a point-like source coincident with the position of RCW~86 was subsequently reported by \citet{RCW86.Fermi.2014}.
With 6.5 yr data, a study performed by \citet{RCW86.Fermi.2016} showed that the \gray\ emission of RCW~86 probably originates from a spatially extended ring region with a hard GeV spectrum with a photon index of $\Gamma\sim1.42$.
The spectrum extends into the TeV range with a high-energy cutoff as revealed by H.E.S.S. observations \citep{RCW86.HESS.2009,RCW86.HESS.2018}.
For $\Delta\alpha=0.8$ and $\beta=2.0$, we obtain the following best-fit model parameters: $\alpha=2.2$, $W_e=9.0\times10^{47}$ erg, $B_{\rm SNR}=26\ \uG$, $\Ebre=2.5$ TeV, and $\Ecut=50$ TeV.

\subsection{RX~J0852.0$-$4622 (G266.2$-$1.2)}
% RX~J0852.0$-$4622 (G266.2$-$1.2) \quad
SNR RX~J0852.0$-$4622, also known as Vela Jr., overlapping the southeast corner of the Vela SNR, was discovered in X-rays by {\sl ROSAT} satellite \citep{Aschenbach1998}.
It has faint radio emission with a spectral index of $\sim$ 0.3 and a rim-brightened morphology similar to that in the X-rays \citep{Combi1999}. 
Like radio and X-ray images, the TeV \gray\ emission also has an almost circular shape with rim enhancement, namely, showing the shell structure \citep{J0852.HESS.2007}.
A detailed analysis of the \hess\ data \citep{J0852.HESS.2018} shows that the TeV \gray\ spectrum can smoothly connect to the hard spectrum in the GeV band detected by \Fermi-LAT \citep{J0852.Fermi.2011}.
In combination with the lack of thermal X-ray emission \citep[e.g.,][]{Slane2001}, these observations strongly favor a leptonic origin for the \gray\ emission. 
For $\Delta\alpha=0.8$ and $\beta=2.0$, we obtain the following best-fit model parameters: $W_e=6.5\times10^{47}$ erg, $\alpha=2$, $B_{\rm SNR}=9 \uG$, $\Ebre=3.0$ TeV, and $\Ecut=53$ TeV.

\subsection{HESS~J1731$-$347 (G353.6$-$0.7)}%\quad
% HESS~J1731$-$347 (G353.6$-$0.7) \quad
HESS~J1731$-$347 was first detected in the H.E.S.S. survey project \citep{survey.HESS.2006} and was identified as an SNR via analysis of radio, IR, and X-ray data \citep{Tian2008.J1731}. It has a faint radio emission and bright nonthermal X-rays.
The systematic analysis of the \hess\ data performed by \citet{J1731.HESS.2011} showed a shell structure in the TeV band and a power-law spectrum with a high-energy cutoff.
In the GeV band, no emission was found in early studies \citep{J1731.Fermi.2014,Acero2015.shell}, but a point source with hard spectrum toward this remnant was detected by \Fermi-LAT with more accumulated data \citep{ SN1006.J1731.Fermi.2017,J1731.Fermi.2018}.
These observational features are very similar to the young SNR RX~J1713.7$-$3946, but their ages are quiet different.
Assuming a distance of 3.2 kpc, the dynamical age is roughly estimated to be $t\approx2.7\times10^{4}$ yr \citep{Tian2008.J1731}, which is much older than that of RX~J1713.7$-$3946.
It should be noted that the ambient density $n_0\sim5\ {\rm cm^{-3}}$ used in determining the dynamical age is derived from the hadronic scenario for the origin of TeV \gray{s}.  
Similar to SNR RX~J1713.7$-$3946, due to lack of thermal X-rays, the ambient density $n_0<0.01\ {\rm cm^{-3}}$ was derived by \citet{J1731.HESS.2011}, suggesting a young age \citep{Acero2015.shell}.
On the other hand, the distance to this source is also a matter of debate.
Based on association with clouds at a velocity range of $-$90 km s$^{-1}$ to $-$75 km~s$^{-1}$, \citet{Fukuda2014} derived a larger distance of 5.2--6.1 kpc,
although a short distance $d\sim3.2$ kpc is preferred to account for multiband observations \citep{Tian2008.J1731,Klochkov2015,Maxted2018.J1731}.
In this paper, we adopt a distance of $d=3.2$ kpc and a young age of $\sim2.5$ kyr \citep[e.g.,][]{SNR.HESS.2018}, implying a current shock velocity of $u_s\sim2200\ {\rm km\ s^{-1}}$.

For $\Delta\alpha=0.8$ and $\beta=2.0$ and assuming infrared photons with a temperature of 40 K and an energy density of 1.0 eV~cm$^{-3}$ \citep{J1731.HESS.2011}, we obtain $\alpha=1.9$ $W_e=2.8\times10^{47}$ erg, $B_{\rm SNR}=26\ \uG$, $\Ebre=3.0$ TeV, and $\Ecut=30$ TeV from the spectral fit.

\subsection{G150.3$+$4.5}

The eastern segment of G150.3+4.5 was first reported by \citet[][called G150.8+3.8]{Gerbrandt2014}, and was considered a strong SNR candidate due to its semicircular shape, nonthermal radio spectrum, and red optical filaments.
Using Urumqi $\lambda$6cm survey data, \citet{Gao2014.G150} found a western and a northern shell beside the eastern shell, forming an enclosed oval shell structure with size of $2.5^{\circ}\times3.0^{\circ}$
and, combining with Effelsberg 11 and 21cm data and CGPS 1420 MHz and 408 MHz observations, they confirmed it to be an SNR.
Based on the HI observations and the Galactic rotation curve, \citet{Cohen2016.PhDT} obtained three possible distances and suggested the nearest distance of 0.4~kpc, due to its large angular size.
Using this distance and the Sedov evolution model, the age is found to be between 0.5 and 5 kyr \citep{Cohen2016.PhDT}.
Here we adopt 3 kyr as its age and derive a velocity of $\sim$1500 ${\rm km\ps}$.
At high-energy bands, it was spatially coincident with the extended \Fermi-LAT source 2FHL J0431.2+5553e \citep{2FHL.Fermi.2016}.
Based on seven years of Pass 8 data and the \rosat\ data, \citet{Cohen2016.PhDT} performed a detailed analysis and obtained a hard spectrum with a photon index of $\varGamma\approx1.8$ in the 1 GeV to 1 TeV band and an X-ray flux upper limit of $\sim5\times10^{-11}\ {\rm erg\ cm^{-2}\ s^{-1}}$ in the 0.5--2 keV band.
For $\Delta\alpha=0.8$ and $\beta=2.0$, we obtain the following parameters: $\alpha=1.7$ $W_e=0.3\times10^{47}$ erg, $B_{\rm SNR}=4\ \uG$, $\Ebre=0.4$ TeV, and $\Ecut=60$ TeV.
Due to the lack of radio flux measurement of the western and northern shells, we double the radio flux of the eastern shell and treat it as the flux for the whole remnant as shown by the gray dots in Figure~\ref{fig:sed}.

\subsection{HESS~J1534$-$571 (G323.7$-$1.0)}

HESS~J1534$-$571 was discovered in the \hess\ Galactic Plane Survey \citep{J1534.HESS.2015} and has a shell-type structure that matches the radio image of the SNR candidate G323.7$-$1.0, confirming its classification as an SNR \citep{newSNR.HESS.2018}.
Its radio counterpart G323.7$-$1.0 was first discovered in the second epoch Molonglo Galactic Plane Survey at 843 MHz, and shows a faint oval shell with a size of $51'\times38'$ \citep{Green2014.MGPS2}.
Based on the $\Sigma-D$ relation and the CO observations, \citet{Maxted2018.J1534} suggested that this SNR evolved in a wind-blown cavity associated with the Scutum-Crux arm gas at a distance of 3.5 kpc.
At this distance, an age of 8--24 kyr and a velocity of 400--1200 ${\rm km\ps}$ are derived from the Sedov relation \citep{Maxted2018.J1534}.
In this paper, we adopt an age of 10 kyr and a velocity of 1000 ${\rm km\ps}$, which are also consistent with that estimated by scaling an RX J1713.7$-$3946-like SNR in the Sedov phase.
Due to the lack of full measurement of its radio flux, we use 0.49 Jy to be a lower limit and 0.98 Jy to be an upper limit following \citet{Maxted2018.J1534}.
At the X-ray band, based on the analysis of the \suzaku\ data, no significant nonthermal emission was detected toward this SNR, giving an X-ray flux upper limit of $\sim2\times10^{-11}\ {\rm erg\ cm^{-2}\ s^{-1}}$ in the 2--12 keV band \citep{newSNR.HESS.2018}.
At the GeV band, \citet{J1534.Fermi.2017} analyzed 8.5 years of \Fermi-LAT data and found an extended GeV source with a hard spectrum $\Gamma\approx1.34$ inside the shell.
For $\Delta\alpha=0.8$, $\beta=2.0$, and infrared photons with a temperature of 20 K and an energy density of 0.8 eV~cm$^{-3}$ \citep{J1534.Fermi.2017}, we obtain the following best-fit model parameters: $\alpha=1.3$, $W_e=6.0\times10^{47}$ erg, $B_{\rm SNR}=5\ \uG$, $\Ebre=10$ GeV, and $\Ecut=15$ TeV.
For this remnant, the electron index below the break is significantly harder than that of others, which is consistent with the fact that older SNRs trend to have harder radio spectra \citep{Reynolds2012,Zeng2018.snrs},
and may be tested with future radio observations.
Given the lack of constraints from radio and X-ray observations, the SED can certainly be fitted with a simpler single power-law model. For the purpose of developing a unified model below, we adopt these broken power-law results.

%%%%%%%%%%%%%%%%%%%%%%%%%%%%%%%%%%%%%%%%%%%%%%%%%%%%%%%%%%%%%%%%%%%%%%%%%%%%%%%%
%%%%%%%%%%%%%%%%%%%%%%%%%%%%%%%%%%%%%%%%%%%%%%%%%%%%%%%%%%%%%%%%%%%%%%%%%%%%%%%%
%%%%%%%%%%%%%%%%%%%%%%%%%%%%%%%%%%%%%%%%%%%%%%%%%%%%%%%%%%%%%%%%%%%%%%%%%%%%%%%%
\section{Discussion}

In our modeling, the two parameters $\Delta\alpha$ and $\beta$ are fixed artificially for all sources with a broken power-law particle distribution.
% $\beta = 2$ is needed to fit the X-ray spectrum of RX~J1713.7$-$3946.
$\Delta\alpha=0.8$ can give a good fit to spectra of all sources.
Of course, the best-fit value may be different for different sources. For example, $\Delta\alpha=0.6$ for RX~J0852.0$-$4622 and $\Delta\alpha=1.0$ for RX~J1713.7$-$3946 will give better fits.
In this study, however, we try to develop a unified model. 
For better comparison of model parameters, it is necessary to have as few free parameters as possible. 
For SN~1006, a single power-law model is adopted and a model with $\beta = 2$ will lead to a poor fit to the X-ray spectrum. We therefore assume $\beta = 1$.
Moreover, as we will see below, for all sources studied here, the break energy is comparable to the age-limited maximum energy, implying that particle acceleration above the break energy already stops at present and the shape of the cutoff is dictated by the radiative energy loss process. 
$\beta = 2$ gives a cutoff sharp enough to fit the X-ray spectra of RX~J1713.7$-$3946. This is not the case for SN~1006, where the acceleration and energy loss time scales are comparable at the cutoff energy.

\begin{figure}[htp]
\centering
\includegraphics[height=60mm,angle=0, trim=50 0 50 25, clip=true]{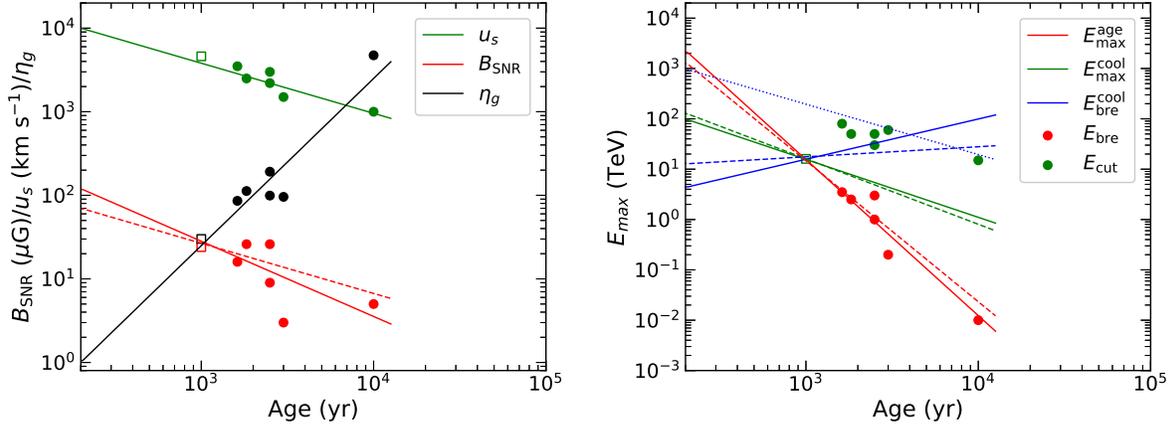}
\caption{Evolution of model parameters with $\tsed=200$~yr, $u_0=10000\km\ps$, $B_0=120\uG$, $\alpha_B=0.9$, and $\lambda=2.0$. The data are listed in Tables~\ref{tab:samp} and \ref{tab:params}. Left: The green, red, and black lines represent the shock velocity, the magnetic field, and the gyrofactor, respectively. Right: the red, green, and blue solid lines show the age-limited maximum energy, the cooling-limited maximum energy and the cooling break energy, respectively. 
The red dots are for the spectral breaks, and the green dots are for the cutoff energies.
The dashed lines in both panels correspond to $B_0=70\uG$ and $\alpha_B=0.6$.
The blue dotted line in the right panel shows the electron energy where the radiative energy loss in an magnetic field of $8\uG$ is equal to the age of the SNRs.
}
\label{fig:evol}
\end{figure}

Because the radiative energy loss time scale at the break energy is much longer than the age of the corresponding SNR, according to the model proposed by \citet{Ohira2012}, this break energy should be associated with the age-limited maximum energy. 
Then, one can obtain $\eta_g$ for each source. The left panel of Figure~\ref{fig:evol} shows the dependence of the shock speed $u_s$, magnetic field $B_{\rm SNR}$, and $\eta_g$ on the age of the SNR. 
Following \citet{Ohira2012}, we divide the evolution of the shock into free expansion and Sedov phases, which starts at $\tsed$, and assume $u_s=u_0(t/\tsed)^{m-1}$, $B_{\rm SNR}=B_0(t/\tsed)^{-\alpha_B}$, $\eta_g=\eta_0(t/\tsed)^{\lambda}$ for $t>\tsed$.
For $\tsed=200$~yr, $u_0=10000\km\ps$, $B_0=120\uG$, $\eta_0=1$, $m=0.4$, $\alpha_B=0.9$, and $\lambda=2.0$, we obtain a good fit to the evolution of $u_s$, $B_{\rm SNR}$, and $\eta_g$.
The latter can be obtained with Equation~\eqref{eq:emax_age} assuming $\Emaxa=\Ebre$.
The corresponding age-limited maximum energy, cooling-limited maximum energy, and cooling break energy are shown with the red, green, and blue solid lines in the right panel of Figure~\ref{fig:evol}, respectively.
The cutoff and break energies of the spectral fits are consistent with such a scenario. 
We therefore have a unified model for the evolution of the high-energy electron distribution accelerated by their shocks. 
The decrease of $\alpha$ and $\Ebre$ with the SNR age $t$ is also consistent with the finding by \citet{Zeng2018.snrs}.
Since SNRs studied here likely have different kinds of progenitors, it is not surprising that parameters of individual sources may have significant deviations from the unified model. G150.3+4.5 and G323.7$-$1.0 appear to be evolving in very low density environments with very low magnetic fields, implying very inefficient particle acceleration.
Moreover, escape processes may play an important role in reducing the energy content of high-energy electrons trapped in relatively order SNRs.

In the Sedov phase, $\alpha_B=0.9$ implies that the magnetic field scales as $B_{\rm SNR}\propto u_s^{3/2}$ which is predicted by some nonlinear diffusive shock acceleration (DSA) theory \citep[e.g.,][]{Bell2004} and is supported by observation data including SN 1993J \citep{Vink2008}.
If the magnetic field amplification is dominated by turbulence stretching in the downstream, the magnetic field should be proportional to $u_s$.
We also plot this case ($\alpha_B=0.6$) with dashed lines in Figure~\ref{fig:evol} with $B_0=70\uG$.

As the youngest SNR in our sample, SN~1006 is different from others for their single power-law distribution.
According to the unified model, this source is going through a particular stage of the shock evolution when the acceleration and energy loss time scale at the cutoff energy is comparable to its age, justifying the single power-law distribution. 
After this stage, the maximum energy to which the shock can accelerate particles is age limited, corresponding to the break energy of our spectral fit.
Electrons above this break energy then only experience radiative energy loss.
Therefore, the maximum energy that the shock of an SNR can accelerate electrons to is comparable to the cutoff energy of SN~1006.
However, the cutoff energies of other SNRs in our sample are higher than those of SN~1006.
This is mainly due to the fact that the $\beta$ for SN~1006 is different from that for others.
For a given high-energy emission, a more gradual cutoff of the electron distribution leads to a lower value for the cutoff energy.
We also notice that for a few sources, the energy loss timescale at the cutoff energy is shorter than the age of the corresponding SNR.
These high-energy electrons likely come from upstream of the shock where the magnetic field is weak and enter the main emission region in the downstream as the remnant expands.
The dotted line in the right panel of Figure~\ref{fig:evol} shows the electron energy where the synchrotron energy loss time scale in a magnetic field of 8 $\uG$ is equal to the age $t$.

From $\alpha_B$ and $\lambda$, one can obtain the time evolution of the break energy $\Emaxa \propto t^{-1+2m-\alpha_B-\lambda} = t^{-3.1}$.
The maximum energy that the shock can accelerate particles to therefore decreases very rapidly as the shock slows down.
Such a sharp decrease is closely related to the decrease of the magnetic field and the increase of $\eta_g$.
Together with the injection rate into the shock acceleration process and the escape rate from SNRs, it determines $\Delta\alpha$.
For $\Delta\alpha = 0.8$, we find that the normalization of electrons below the break energy should scale as $t^{0.8*3.1}=t^{2.5}$, implying that the injection rate into the shock acceleration process should be proportional to $t^{1.5}$ in the absence of escape.
Our spectral fit shows that $W_e$ is more or less independent of $t$ (Table~\ref{tab:params}). 
The particle escape process therefore must play a role in determining $W_e$.
Since $W_e$ is almost independent of $t$, a broken power-law distribution of high-energy electrons may be realized with an energy independent escape time $t_{\rm esc}\propto t$ and an injection rate inversely proportional to $t$. Then the ion injection needs to be different from electrons to give rise to very efficient low-energy ion acceleration in old SNRs \citep{Zhang2018.Acc}.

\section{Conclusion}

By fitting the nonthermal spectra of seven shell-type middle-age \gray\ SNRs with a leptonic model, we obtain the high-energy electron distribution in these remnants and the mean magnetic field of the emission region.
We find that the magnetic field decreases rapidly with the SNR age as $B_{\rm SNR} \propto t^{-0.9}$ and, in general, the electron distribution is consistent with a broken power law with a high-energy cutoff.
The characteristics of the spectral evolution are also consistent with the model proposed by \citet{Ohira2012}, where the maximum energy that shocks in the Sedov phase can accelerate particles to decreases rapidly and the spectral break is associated with the age-limited maximum energy.
In particular, we find that the age-limited maximum energy scales as $t^{-3.1}$.
In combination with the spectral index change $\Delta\alpha = 0.8$, the injection rate in the shock acceleration process scales as $t^{1.5}$.
The rapid decrease of the age-limited maximum energy is closely related to the decrease of the magnetic field and the rapid increase of the diffusion coefficient, with $\eta_g\propto t^{2.0}$.
The rapid increase of the injection rate indicates that most of the low-energy CRs are accelerated in old SNRs, as suggested by \citet{Zhang2017.LY}.
This study needs to be extended to include young SNRs in order to have a complete picture of particle acceleration in SNRs.
A two-zone emission model may be needed to account for prominent nonthermal emissions associated with reverse/inward shocks of young SNRs \citep{Zhang2019.Casa}.

\section*{Acknowledgements}
We appreciate helpful discussions with Xiaoyuan Huang and Qiang Yuan on \Fermi\ data.
X.Z. is indebted to Yang Chen for the helpful discussion on the evolution of electron spectrum.
This work is partially supported by National Key R\&D Program of China 2018YFA0404203 and 2018YFA0404204, NSFC grants U1738122, 11761131007, and by the International Partnership Program of Chinese Academy of Sciences, grant No. 114332KYSB20170008.
X.Z. also thanks the support of NSFC grants 11803011, 11851305 and 11633007.

\bibliographystyle{aasjournal}

\bibliography{/Users/xiao/papers/bib/refs,/Users/xiao/papers/bib/refs-book,/Users/xiao/papers/bib/refs-G-ray,/Users/xiao/papers/bib/refs-obs-HE}

%%% version 3
\begin{deluxetable}{p{6pc}ccccccc}
\tabletypesize{\scriptsize}
\tablecaption{References and physical parameters for our sample.
\label{tab:samp}}
\tablewidth{0pt}
\tablehead{
\colhead{Source} &\colhead{Age} & \colhead{Distance} & \colhead{Velocity} & 
\colhead{$\alpha_{r}$\tablenotemark{a}} & Radius\tablenotemark{b}& Radius\tablenotemark{c} & \colhead{References} \\
 & \colhead{(kyr)} &  \colhead{(kpc)} & \colhead{(km s$^{-1}$)}&  & (pc) & (pc) & 
}
\startdata
 SN~1006           & 1.00  & 2.2  & 4600   & 0.6   &  9.6 &  9.8 & 1--3 \\
 RX~J1713.7$-$3946 & 1.62  & 1.0  & 3500   & ?     &  8.7 & 11.8 & 4--6 \\
 RCW~86            & 1.83  & 2.5  & 2500   & 0.6   & 15.3 & 12.4 & 7--11   \\
 RX~J0852.0$-$4622 & 2.50  & 0.75 & 3000   & 0.3\ ?& 13.1 & 14.0 & 12  \\
 HESS~J1731$-$347  & 2.50  & 3.2  & 2200   & 0.4   & 14.0 & 14.0 & 13--15    \\
 G150.3+4.5        & 3.00  & 0.4  & 1500   & ?     & 10.5 & 15.1 & 16 \\
 HESS~J1534$-$571  & 10.00 & 3.5  & 1000   & ?     & 22.9 & 24.4 & 17
\enddata
\tablecomments{
(1) \citealt{Winkler2003}, (2) \citealt{Katsuda2009.SN1006}, (3) \citealt{Winkler2014}; 
(4) \citealt{Wang1997.QC}, (5) \citealt{Fukui2003}, (6) \citealt{Zirakashvili2010};
(7) \citealt{Rosado1996}, (8) \citealt{Sollerman2003}, (9) \citealt{Vink2006}, (10) \citealt{Helder2013}, (11) \citealt{Yamaguchi2016};
(12) \citealt{Katsuda2008};
(13) \citealt{Tian2008.J1731}; (14) \citealt{J1731.HESS.2011}; (15) \citealt{SNR.HESS.2018};
(16) \citealt{Cohen2016.PhDT};
(17) \citealt{Maxted2018.J1534}.
}
\tablenotetext{a}{Radio spectral index.}
\tablenotetext{b}{Calculated from the angular size and the distance.}
\tablenotetext{c}{Calculated from $R=2.5u_0t_{\rm sed}(t/t_{\rm sed})^{2/5}$.}
\end{deluxetable}

\begin{deluxetable}{lccccccccccc}
\tabletypesize{\scriptsize}
\tablecaption{Fitting parameters.\label{tab:params}}
\tablewidth{0pt}
\tablehead
{
    \colhead{source} & \colhead{$\beta$} & \colhead{$\Delta\alpha$} & \colhead{$\alpha$} & \colhead{$\Ebre$} & \colhead{$\Ecut$} & \colhead{$W_e$} & \colhead{$B_{\rm SNR}$} & $\tau_{syn}(\Ecut)$ &\colhead{$\eta_g^{a}$}\\
       &              &             &       & \colhead{(TeV)}         & \colhead{(TeV)}         & \colhead{($10^{47}$ erg)} & \colhead{(${\rm \mu G}$)} &  (kyr) &
}
\startdata
SN 1006           & 1.0 & ---  & 2.2  & ---    & 16 & 1.7 & 24  & 1.4 & $30$  \\
RX J1713.7$-$3946 & 2.0 & 0.8  & 2.0  & 3.5    & 80 & 4.0 & 16  & 0.6 & $86$  \\
RCW 86            & 2.0 & 0.8  & 2.2  & 2.5    & 50 & 9.0 & 26  & 0.4 & $113$ \\
RX J0852.0$-$4622 & 2.0 & 0.8  & 1.9  & 1.0    & 50 & 8.0 & 9   & 3.1 & $192$  \\
HESS J1731$-$347  & 2.0 & 0.8  & 1.9  & 3.0    & 30 & 2.8 & 26  & 0.6 & $99$  \\
G150.3+4.5        & 2.0 & 0.8  & 1.7  & 0.2    & 60 & 0.4 & 3   & 23.2& $96$ \\
HESS J1534$-$571  & 2.0 & 0.8  & 1.3  & 1.0E$-2$ & 15 & 6.0 & 5   & 33.5& $4731$
\enddata
\tablenotetext{a}{Calculated from $\Emaxa=\Ebre$.}
\end{deluxetable}

\end{document}